\documentclass[twocolumn,prl,showpacs,amsmath,amssymb]{revtex4}
\usepackage[latin1]{inputenc}
\usepackage{graphicx}
\usepackage{dcolumn}
\usepackage{bm}

\begin{document}



\title{Magnetic dipolar ordering and quantum phase transition in Fe$_{8}$ molecular magnet.}

\author{E. Burzur\'{\i}}
\affiliation{Instituto de Ciencia de
Materiales de Arag\'on, C.S.I.C. - Universidad de Zaragoza, and Dpto.
de F\'{\i}sica de la Materia Condensada, Universidad de Zaragoza, E-50009 Zaragoza, Spain}

\author{F. Luis}
\email{fluis@unizar.es} \affiliation{Instituto de Ciencia de
Materiales de Arag\'on, C.S.I.C. - Universidad de Zaragoza, and Dpto.
de F\'{\i}sica de la Materia Condensada, Universidad de Zaragoza, E-50009 Zaragoza, Spain}%

\author{O. Montero}
\affiliation{Instituto de Ciencia de
Materiales de Arag\'on, C.S.I.C. - Universidad de Zaragoza, and Dpto.
de F\'{\i}sica de la Materia Condensada, Universidad de Zaragoza, E-50009 Zaragoza, Spain}%

\author{J. Campo}
\affiliation{Instituto de Ciencia de
Materiales de Arag\'on, C.S.I.C. - Universidad de Zaragoza, and Dpto.
de F\'{\i}sica de la Materia Condensada, Universidad de Zaragoza, E-50009 Zaragoza, Spain}%

\author{B. Barbara}
\affiliation{Institut N\'eel, CNRS \& Universit\'e Joseph Fourier, BP166, 38042 Grenoble Cedex 9, France}

\author{R. Ballou}
\affiliation{Institut N\'eel, CNRS \& Universit\'e Joseph Fourier, BP166, 38042 Grenoble Cedex 9, France}

\author{E. Ressouche}
\affiliation{CEA/Grenoble, INAC/SPSMS-MDN,17 rue des Martyrs, 38054 Grenoble Cedex 9, France}

\author{S. Maegawa}
\affiliation{Graduate School of Human and Environmental Studies, Kyoto University, Kyoto 606-8501, Japan}


\date{\today}


\begin{abstract}

We show that a crystal of mesoscopic Fe$_{8}$ single molecule magnets is an experimental realization of the Quantum Ising Phase Transition (QIPT) model in a transverse field, with dipolar interactions. Quantum annealing has enabled us to explore the QIPT at thermodynamical equilibrium. The phase diagram and critical exponents we obtain are compared to expectations for the mean-field QIPT Universality class.

\end{abstract}


\maketitle



Quantum phase transitions \cite{Hertz76,Sachdev99} (QPT) have been extensively studied in recent years. Physical realizations include the superconductor insulator canonical phase transition in cuprates \cite{Sachdev92,Chubukov93,Sokol93}, the onset of antiferromagnetism in heavy fermions \cite{Barbara89,Barbara92,Mathur98,Gegenwart08,Si10}, the pressure driven insulator-metal transition in V$_{2}$O$_{3}$ \cite{Carter91}, and the order-disorder magnetic transitions driven by field (LiHoYF$_{4}$ \cite{Bitko96}) or concentration (Cr$_{x}$V$_{1-x}$ alloys \cite{Yeh02}). In addition to their intrinsic interest, a plethora of new properties arise at nonzero temperature.

In magnetism, the QIPT model is realized by a lattice of $N$ coupled Ising spins in a transverse magnetic field $H_{\perp}$ \cite{Sachdev99}, the Hamiltonian of which reads as follows

\begin{equation}
{\cal{H}} = -2S^2\sum_{i<j}^{N}J_{ij}\sigma_{i}^{z}\sigma_{j}^{z}-\Delta_{S}\sum_{i}^{N}\sigma_{i}^{x}
\label{Ising}
\end{equation}

\noindent Here, $\sigma'$s are the Pauli spin operators, $J_{ij}$ the longitudinal couplings and $\Delta_{S}$ the ground-state tunnel splitting which depends on and vanishes with $H_{\perp}$. The classical long-range order that exists for $H_{\perp} = \Delta_{S} = 0$ (say, ferro- or antiferro-magnetic) competes with the field-induced quantum fluctuations ($[{\cal{H}},\sigma] \neq 0$). Long-range order is completely destroyed at the QIPT critical point when $\Delta_{S}(H_{\perp}) \gtrsim \Delta_{S{\rm c}}$ where $\Delta_{S {\rm c}} = 2JS^{2} \sim k_{\rm B}T_{\rm c}$ and $J = (1/N)\sum_{i<j} J_{ij}$. The ground-state becomes then a superposition of "up" and "down" spin states (here $T_{\rm c}$ is the Curie temperature of a ferromagnet, which is the case for Fe$_{8}$).

Crystals of interacting dipoles are natural candidates to observe a QPT \cite{Bitko96}. Because of the weak interaction strength, the effects of quantum fluctuations can become noticeable at moderate transverse fields. In crystals of single-molecule magnets (SMMs), this effect is amplified because of very large inter-molecular distances \cite{Morello03}. Furthermore, hyperfine interactions in $3d$-based SMMs are generally weak enough not to block the QPT, whereas they play an important role in the case of the best known lanthanide-based insulators \cite{Aeppli05}. Finally, SMMs show quantum phenomena like tunneling \cite{Barbara95,Friedman96,Hernandez96,Thomas96,Sangregorio97}, interference\cite{Wernsdorfer99} and superpositions of spin states \cite{Luis00}. However, their complex mesoscopic character suggests that quantum spin fluctuations and therefore the QPT may, in these systems, be affected by decoherence \cite{Bertaina08}, therefore providing an interesting playground to investigate QPT in the presence of a noisy environment \cite{DallaTorre10}.

A first indication of a possible QPT in a molecular crystal of Mn$_{12}$ acetate SMM was inferred from magnetic neutron diffraction experiments \cite{Luis05}. However, the experimental uncertainties close to $M \simeq 0$ hindered the exploration of the critical region. In addition, the local anisotropy axes of Mn$_{12}$ acetate molecules are slightly tilted due to the presence of several isomers giving rise, in the presence of a perpendicular field, to random fields. Very recently \cite{Wen10}, it has been argued that such fields turn Mn$_{12}$ acetate into a realization of the classical random-field Ising model. The existence of a pure QPT in a crystal of molecular nanomagnets is, therefore, not yet established. Our aim here is to explore the experimental realization of the QIPT model in a disorder-free SMM crystal.

We have chosen as a candidate the [(C$_{6}$H$_{15}$N$_{3}$)$_{6}$Fe$_{8}$O$_
{2}$(OH)$_{12}$] system \cite{Barra96}, hereafter referred to as Fe$_{8}$. Each molecule has a net spin $S=10$ and a strong uniaxial magnetic anisotropy, which can be described by the following spin hamiltonian

\begin{equation}
{\cal H}_{0} = -DS_{z}^{2} + E \left(S_{x}^{2}-S_{y}^{2} \right)
\label{Hamiltonian0}
\end{equation}

\noindent where $D/k_{\rm B} = 0.294$ K and $E/k_{\rm B} = 0.046$ K. Equation (\ref{Hamiltonian0}) defines $x$, $y$ and $z$ as the hard, medium and easy magnetization axes. In the triclinic structure of Fe$_{8}$, $x$, $y$ and $z$ axes are common to all molecules \cite{Ueda01}.

Susceptibility experiments were performed on a single crystal \cite{Ueda01} of approximate dimensions $3 \times 2 \times 1$ mm$^{3}$ and $m = 4.72$ mg. The complex magnetic susceptibility $\chi = \chi^{\prime}(T,\omega) - i \chi^{\prime \prime}(T,\omega)$ was measured down to $90$ mK and in the frequency range $3$ Hz $\leq \omega/2 \pi \leq 20$ kHz using a homemade ac susceptometer thermally anchored to the mixing chamber of a $^{3}$He-$^{4}$He dilution refrigerator. A dc magnetic field $\overrightarrow{H}$ was applied with a $9$ T$\times 1$ T$\times 1$ T superconducting vector magnet, which enables rotating $\overrightarrow{H}$  with an accuracy better than $0.001^{\circ}$.

The orientation of the magnetic anisotropy axes \cite{Ueda01} was first approximately determined through the identification of the crystallographic axes by X-ray diffraction. The magnetic easy axis $z$ was oriented parallel to the ac excitation magnetic field of amplitude $h_{\rm ac} = 0.01$ Oe. The sample was completely covered by a non-magnetic epoxy to prevent it from moving under the action of the applied magnetic field. The alignment of $\overrightarrow{H}$ perpendicular ($\pm 0.05 ^{\circ}$) to $z$  and close ($\pm 20 ^{\circ}$) to the medium $y$ axis was done at low temperatures ($T = 2$ K), using the strong dependence of $\chi^{\prime} (T,\omega)$ on the magnetic field orientation.

Neutron diffraction experiments were performed on the thermal neutron two-axis diffractometer for single crystals D23 at the Institute Laue Langevin. A single crystal, thermally anchored to the mixing chamber of a a $^{3}$He-$^{4}$He dilution refrigerator ($T \geq 50$ mK), was mounted so as to have the anisotropy axis orthogonal to the magnetic field. Longitudinal field components arising from the nonperfect orientation of the crystal were compensated by means of a superconducting mini-coil inserted in the space between the dilution fridge and the superconducting magnet. The intensities of different Bragg peaks, in particular $(010)$ to which the magnetization components $M_{z}$ and $M_{x}$ strongly contribute with roughly equal weights and $(-100)$ or $(-2 1 2)$ to which $M_{y}$ and $M_{x}$ contribute significantly, were measured as a function of $T$ (under a constant magnetic field) and $H_{\perp}$ (for a constant temperature). The nonmagnetic contributions were measured at high temperature ($5$ K) and $H_{\perp} = 0$. The fit of the magnetic intensities enabled us to estimate the magnetization components along the external magnetic field $M_{\perp}$ and along the anisotropy axis $M_{z}$.

\begin{figure}[h]
\resizebox{7.5cm}{!}{\includegraphics{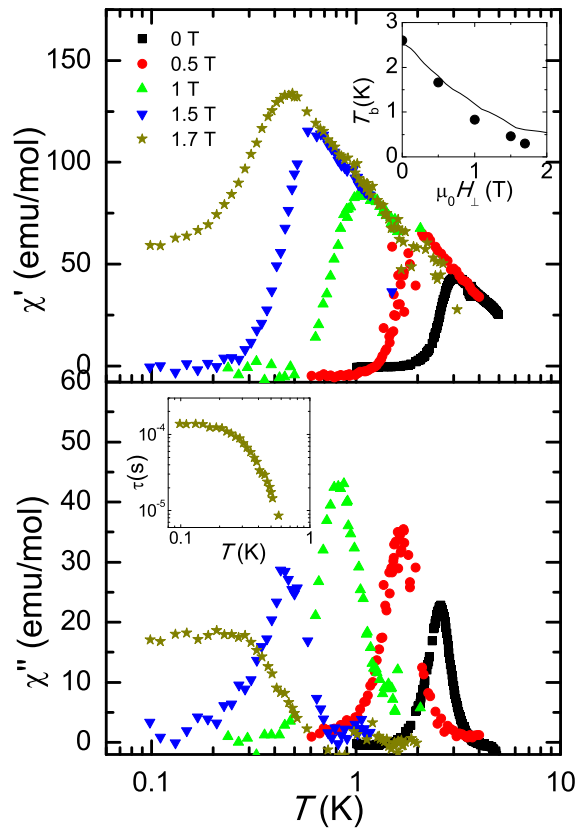}}
\caption{$\chi^{\prime}$ (top) and $\chi^{\prime \prime}$ (bottom) measured at $\omega/2\pi = 333$ Hz and different $H_{\perp}$. The top inset shows the shift of $\chi^{\prime \prime}$ maxima with increasing $H_{\perp}$. The solid line shows theoretical predictions for quantum spin-phonon relaxation that follow from Pauli's master equation as described in \cite{Luis98}. The bottom inset shows the magnetic relaxation time at $\mu_{0}H_{\perp} = 1.7$ T. }
\label{XvsTunderBperp}
\end{figure}

The ac-susceptibility (Figure \ref{XvsTunderBperp}) deviates from equilibrium for low $H_{\perp}$ and low $T$, as shown by the vanishing of $\chi^{\prime}$ and by the relatively large values of $\chi^{\prime \prime}$. Furthermore, the superparamagnetic blocking temperature $T_{\rm b}$ (associated with the $\chi^{\prime \prime}$ maximum) is strongly frequency-dependent (not shown). In these dipolar Ising-like systems, out of equilibrium conditions generally arise from the presence of energy barriers larger than the typical interaction energies (i.e. $T_{\rm b} > T_{\rm c}$), a situation which can be inverted by a quantum annealing process, via the application of a transverse magnetic field \cite{Kadowaki98,Brooke99}. Indeed, when $H_{\perp}$ increases the height of the energy barrier decreases and spins are able to tunnel via lower lying energy levels, which leads to enhanced spin-lattice relaxation rates. This explains qualitatively the decrease of $T_{\rm b}$ towards lower temperatures that we observe, as shown in the inset of Fig. \ref{XvsTunderBperp}.

\begin{figure}[h]
\resizebox{7.5cm}{!}{\includegraphics{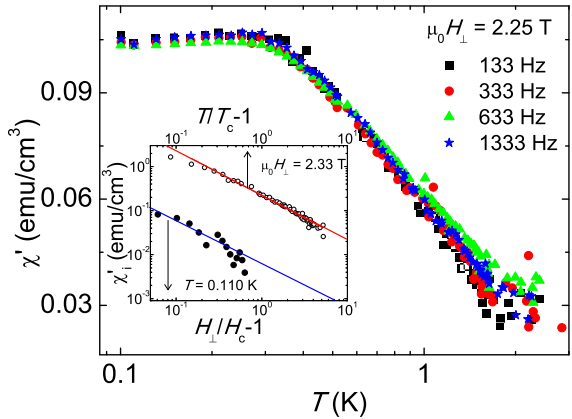}}
\caption{ $\chi^{\prime}$ measured at $\mu_{0} H_{\perp} = 2.25$ T. The data are independent of frequency for any $T>0$. The inset shows a log-log plot of demagnetization-corrected $\chi_{\rm i}^{\prime}$ vs the reduced temperature (for $\mu_{0}H_{\perp}=2.33$ T with $T_{c}=340$ mK, $\circ$) and field (at $T=0.110$ K with $\mu_{0} H_{c}= 2.6$ T, $\bullet$). The slopes of the linear fits give the critical exponents $\gamma_{\rm cl} \simeq 1.02(2)$ and $\gamma_{\rm qu} \simeq 1.1(1)$ .}
\label{order}
\end{figure}

At moderate $H_{\perp}$, $\chi^{\prime}$ is not fully out of equilibrium (e.g. at $\mu_{0} H_{\perp} = 1.7$ T where it remains finite down to lowest temperatures).The relaxation time $\tau = \chi^{\prime \prime}/ \chi^{\prime} \omega$ estimated from these data is already very short and, more importantly, it remains constant below approximately $250$ mK. $\chi^{\prime}$ reaches full equilibrium for $\mu_{0} H_{\perp} \gtrsim 2$ T as seen in Fig. \ref{order} where $\chi^{\prime}$ measured at $H_{\perp} = 2.25$ T becomes independent of frequency and temperature below $340$ mK (which we take as the critical temperature $T_{\rm c}$ at this field). The saturation value corresponds to the inverse demagnetizing factor of our sample $1/N = 0.1$ emu/cm$^{3}$Oe, as expected for an equilibrium ferromagnetic phase transition in which magnetic interfaces can move freely. Here, equilibrium arises because $\Delta_{S}$ is large enough (or/and the barrier small enough) to enable spins to relax via ground-state tunneling.

\begin{figure}[h]
\resizebox{7cm}{!}{\includegraphics{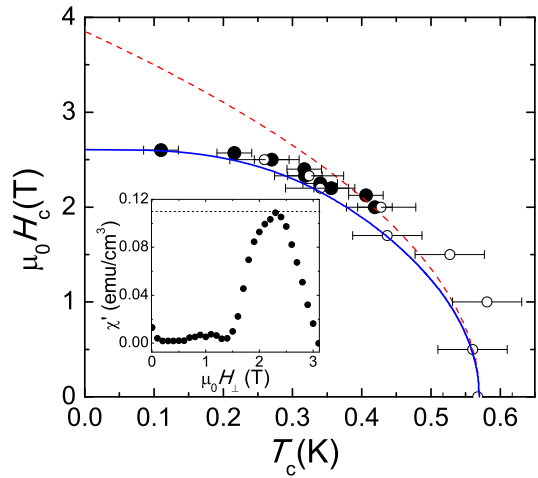}}
\caption{$H_{\rm c}$-$T_{\rm c}$ phase diagram. Solid symbols were obtained from the divergence of the equilibrium intrinsic susceptibility $\chi_{\rm i}$. Open symbols were determined from the Curie-Weiss fit of $1/\chi_{\rm i}$ data measured in the paramagnetic region. The solid line represents a mean-field calculation of the phase boundary using Eq. (\ref{HamiltonianMF}). The dashed line is the classical phase diagram, calculated from Monte Carlo simulations. Inset: $\chi^{\prime}$ vs $H_{\bot}$ measured at $T = 0.110$ K and $\nu = 333$ Hz.}
\label{phasediagram}
\end{figure}

We next show the existence of an equilibrium QPT and discuss its critical behavior. The intrinsic $\chi_{\rm i}^{\prime}$ corrected from demagnetizing effects is plotted vs reduced temperature, at $2.33$ T, and field, at $110$ mK, in the inset of Fig. \ref{order}. Under these conditions, $\chi_{\rm i}^{\prime}$ should follow, as it approximately does, the power laws

\begin{equation}
\chi_{\rm i}^{\prime} = \left( \frac{T-T_{\rm c}}{T_{\rm c}} \right)^{-\gamma_{\rm cl}}, \chi_{\rm i}^{\prime} = \left( \frac{H_{\perp} - H_{\perp {\rm , c}}}{H_{\perp {\rm , c}}} \right)^{-\gamma_{\rm qu}}
\label{gamma}
\end{equation}

\noindent where $T_{\rm c}=340$ mK and $\mu_{0} H_{\rm c} = 2.6$ T are determined experimentally. The slopes give critical exponents $\gamma_{\rm cl} = 1.02(2)$ and $\gamma_{\rm qu} = 1.1 (1)$, in good agreement with $\gamma = 1$ of the Ising mean-field universality class in $3$ dimensions. The $H_{\rm c}$-$T_{c}$ phase diagram shown in Fig \ref{phasediagram} is obtained, above $2$ T, by repeating the procedure described in Fig. \ref{order} for different $H_{\perp}$ and, below $2$ T, from the fit of $(\chi_{i}^{\prime})^{-1}$ to a Curie-Weiss law using the mean-field character of the phase transition (i.e. that $\gamma_{\rm cl} = 1$ and that the Weiss temperature $\theta \simeq T_{\rm c}$).

As expected, $T_{\rm c}$ decreases when quantum fluctuations increase (with $H_{\perp}$) showing a phase transition induced by thermal fluctuations at $T_{c}(H_{\perp}=0) = 0.57(5)$ K (semi-classical) and a phase transition induced by quantum fluctuations at $\mu_{0} H_{\rm c}(T = 0) \simeq 2.6$ T (quantum). The observed Curie temperature of $0.57(5)$ K agrees very well with the Monte-Carlo prediction $T_{\rm c} = 0.54$ K \cite{Fernandez00}. The solid line is obtained by solving numerically the mean-field Hamiltonian for the $S=10$ spin

\begin{equation}
{\cal H} = {\cal H}_{0} -g\mu_{\rm B} H_{\perp} (S_{x} \cos \phi  + S_{y} \sin \phi ) - 2 J \langle S_{z} \rangle_{T} S_{z}
\label{HamiltonianMF}
\end{equation}

\noindent where $\phi$ is the angle between the magnetic field and the hard axis, $2J$ is the molecular field coefficient and $\langle S_{z} \rangle_{T}$ is the thermal average of $S_{z}$. We set $J/k_{\rm B} = T_{\rm c}/2S^{2} = 2.85 \times 10^{-3}$ K and $\phi = 70 ^{\circ}$ that accounts for $\mu_{0} H_{\rm c}(T = 0) \simeq 2.6$ T. The agreement between the measured and calculated phase diagrams (Fig. \ref{phasediagram}) is excellent despite that the field range where ferromagnetism is observed at equilibrium (when $\chi^{\prime}$ = $1/N$) can be very narrow as shown in the inset of Fig. \ref{phasediagram}. The dashed line represents the results of classical Monte Carlo simulations \cite{Fernandez10} which agree with classical mean-field predictions. The classical phase boundary is well approximated by $H_{\rm c} (T_{\rm c}) = H_{\rm c} (0) \left[ 1- T_{\rm c}/T_{c}(H_{\perp}=0)\right]^{1/2}$. In this model $H_{\rm c} (0)$ equals the anisotropy field $H_{K} = 2 \left[ D - E \left( \sin^{2} \phi - \cos^{2} \phi \right) \right]/g \mu_{\rm B} S  \simeq 3.8$ T, which clearly overestimates the experimental critical field due to the absence of quantum fluctuations.

\begin{figure}[h]
\resizebox{7cm}{!}{\includegraphics{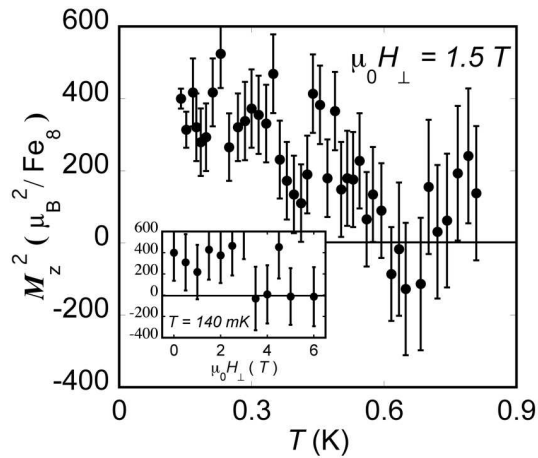}}
\caption{ Squared longitudinal magnetization of a Fe$_{8}$ crystal determined from neutron diffraction intensities measured under a constant $\mu_{0} H_{\perp} = 1.5$ T. The inset shows data measured at a fixed temperature $T=0.14$ K as a function of $H_{\perp}$.}
\label{MLvsT&F}
\end{figure}

Additional evidence supporting the existence of a phase transition to a dipolar ferromagnetic phase can be found in the results of neutron diffraction experiments. Figure \ref{MLvsT&F} shows $M_{z}^{2}$ determined directly from magnetic diffraction intensities measured at $\mu_{0}H_{\perp} = 1.5$ T. The inset shows field-dependent data measured at a fixed temperature of $0.14$ K. Despite the high noise level, typical of these experiments, the results show the onset of a net spontaneous $M_{z}$ in, respectively, the low-$T$ ($\leq 0.6(1)$ K) and low-$H_{\perp}$ ($\leq 3(1)$ T) regions.

Summarizing, we have shown that the single molecular magnet Fe$_{8}$ undergoes a dipolar ferromagnetic to paramagnetic phase-transition of (i) classical nature ($\mu_{0} H_{\rm c} = 0$ T and $T_{\rm c} = 570$ mK) and (ii) quantum nature ($T_{\rm c} = 110$ mK and $\mu_{0} H_{\rm c} = 2.6$ T). Long-range magnetic order between these mesoscopic SMM was enabled by the use of a quantum annealing procedure, which allows the spin system to attain thermal equilibrium despite the presence of high anisotropy energy barriers preventing spin-reversal. While the observed $T_{\rm c}$ agrees very well with classical Monte Carlo calculations \cite{Fernandez00}, the $H_{\rm c}-T_{\rm c}$ phase diagram shows important quantitative and qualitative deviations from the expected classical behaviour at sufficiently low-$T$ (or high-$H_{\perp}$). By contrast, it is in excellent agreement with quantum predictions obtained via the diagonalisation of the mean-field Hamiltonian for the coupled $S=10$ spins. The experiments show that the critical behavior and the phase diagram fulfill the expectations of the 3D Mean-Field Quantum Ising universality class. This agreement allows us to extend the notion of ``macroscopic order parameter" \cite{Leggett00} of SMMs to the ferromagnetic region.


\begin{acknowledgments}
We are grateful to Dr Julio Fern\'andez for useful suggestions and for making available to us the results of his classical Monte Carlo and mean-field simulations. We acknowledge the assistance of Eric Bourgeat-Lami, Xavier Tonon, Pascal Fouilloux, and Bruno Vettard in designing, fabricating, and installing the superconducting mini-coil used in the neutron diffraction experiments, and Tomoaki Yamasaki and Dr. Miki Ueda for the synthesis of the samples. The present work was partly funded by grant MAT2009-13977-C03 (MOLCHIP) and the Consolider-Ingenio project on Molecular Nanoscience, from the Spanish MICINN.
\end{acknowledgments}



\end{document}